\documentclass[aps,pra,twocolumn,superscriptaddress,showpacs,floatfix,nofootinbib]{revtex4}
\usepackage{graphicx}
\usepackage{epsfig}

\newcommand{\nc}{\newcommand}
\nc{\rnc}{\renewcommand}
\nc{\smfrac}[2]{\mbox{$\frac{#1}{#2}$}}
\nc{\eq}[1]{(\ref{#1})}
\def\s{\sigma}
\nc{\w}{\omega}
\nc{\g}{\gamma}
\def\<{\langle}
\def\>{\rangle}
\nc{\ox}{\otimes}
\nc{\be}{\begin{equation}}
\nc{\ee}{{\end{equation}}}
\nc{\bea}{\begin{eqnarray}}
\nc{\eea}{\end{eqnarray}}
\def\sq12{\smfrac{-i}{\sqrt{2}}}
\def\p5{\smfrac{1}{2}}
\def\ps{\vspace*{0.5ex}}
\def\sh{\smfrac{1}{\sqrt{2}}}

\begin{document}

\title{An exact effective two-qubit gate in a chain of three spins}

\date{\today}

\author{Man-Hong Yung}
\email[email: ]{ mhyung@phy.cuhk.edu.hk}

\affiliation{Physics Department, The Chinese University of Hong
Kong, Hong Kong}

\affiliation{Institute for Quantum Information, MSC 107-81, California
Institute of Technology, Pasadena, CA 91125, USA}

\author{Debbie W. Leung}
\email[email: ]{wcleung@caltech.edu}

\affiliation{Institute for Quantum Information, MSC 107-81, California
Institute of Technology, Pasadena, CA 91125, USA}

\author{Sougato Bose}
\email[email: ]{sougato.bose@qubit.org}

\affiliation{Institute for Quantum Information, MSC 107-81, California
Institute of Technology, Pasadena, CA 91125, USA}

\affiliation{Department of Physics and Astronomy, University College
London, Gower Street, London WC1E 6BT}


\begin{abstract}
  We show that an effective two-qubit gate can be obtained from the
  free evolution of three spins in a chain with nearest neighbor XY
  coupling, without local manipulations.  This gate acts on the two
  remote spins and leaves the mediating spin unchanged. It can be used
  to perfectly transfer an arbitrary quantum state from the first spin
  to the last spin or to simultaneously communicate one classical bit
  in each direction.  One ebit can be generated in half of the time
  for state transfer.
  For longer spin chains, we present methods to create or transfer
  entanglement between the two end spins in half of the time required for
  quantum state transfer, given tunable coupling strength and local
  magnetic field.  We also examine imperfect state transfer through a
  homogeneous XY chain.
\end{abstract}

\maketitle

\raggedbottom

\section{Introduction}
One of the most important resources in quantum information processing
is the interaction between different quantum systems, such as those
possessed by remote parties or registers of a quantum computer.
These systems evolve jointly according to some interaction
Hamiltonian.
Good methods to convert such interaction into other useful
resources, using a well motivated set of control operations, are of
paramount importance.

\ps

A common framework for studying such conversions involves a system of
$n$ qubits (or higher dimensional registers) with {\em unlimited local
control}.  Sophisticated methods for various tasks were developed.
For example, in a system coupled by {\em simultaneous} arbitrary
pairwise interactions, one can extract the interaction on a select
pair of qubits \cite{decoupling}, which can be used to
provide a coupling gate or entanglement.
These methods typically approximate the desired task by interspersing
the free Hamiltonian evolution with local operations.  The required
amount of local resources can be reflected by the frequency of
manipulations, and an error parameter is given by the interval between
them.  In general, the task is approximated to first order of this
time interval, though sometimes the exact task can be performed
without excessive local resources
(for examples, see \cite{Leung99,Jones99} for logic gate construction,
and \cite{Bose03,Subrahmanyam03,Christandl03} for state transfer and
entanglement generation).

\ps

In this paper, we consider a simple but nontrivial system in which
several tasks can be accomplished perfectly almost without local
manipulations.
We consider a spin chain with nearest neighbor coupling; the two end
spins are data qubits and other spins are mediating the indirect
coupling between the data qubits.  Furthermore, we restrict local
control to the data qubits only.
The Hamiltonian is given by
\begin{equation}\label{Hgen}
\!\!\!\!  H = \sum_{\<i,j\>} \omega_{ij} \left( {\sigma _ {\!+\,i} \,
\sigma _ {\!-\,j} + \sigma _ {\!-\,i} \, \sigma _ {\!+\,j} } \right),
\end{equation}
where
$\sigma_+ = \left(
\begin{array}{cc}
0 & 0 \\
1 & 0
\end{array}\right)$ and
$\sigma_- = \left( \begin{array}{cc}
0 & 1 \\
0 & 0
\end{array}\right)$
are the raising and lowering operators respectively, $i,j$ are spin
labels; subscripts of operators indicate the spins being acted on,
and the sum is over neighboring spin-pairs denoted by $\<i,j\>$.
Expressing $\sigma_\pm$ in terms of the Pauli matrices
$\{\s_{x,y,z}\}$, we have $\sigma_\pm = \smfrac{1}{2} (\s_x \pm i
\s_y)$ and
\begin{equation}\label{Hgenv2}
\!\!\!\!  H = \sum_{\<i,j\>} \frac{\omega_{ij}}{2} \left( {\sigma _
{\!x\,i} \, \sigma _ {\!x\,j} + \sigma _ {\!y\,i} \, \sigma _ {\!y\,j}
} \right).
\end{equation}
\begin{figure}[http]
\begin{center}
\includegraphics[width=6 cm]{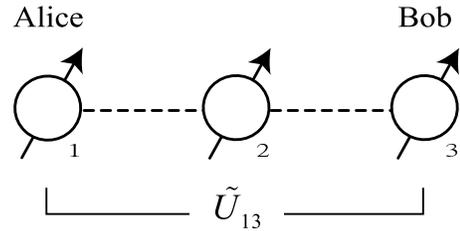}
\caption{A spin chain of three spin particles are interacting under
the Hamiltonian in (\ref{H2}).  An effective two-qubit gate, $\tilde
U_{13}$, can be obtained at $t = \smfrac{\pi}{\sqrt 2 \omega}$, when
$\omega = \lambda$. The gate can perform perfect state transfer from
spin 1 to 3 and also entangle spin 1 with spin 3, leaving spin 2
unentangled.}
\label{3spins}
\end{center}
\end{figure}
Our main result is for the case of a $3$-spin chain
(Fig.\ref{3spins}).  We show how to turn the free evolution into an
{\em exact} effective quantum gate acting on the two data qubits
without affecting the mediating spin.
This gate can be used to transfer an unknown quantum state between the
data spins, to create a maximally entangled state, and to communicate
classical information.
We also explore applications and extensions to this simple basic
result.

\subsection{Related works}

Before presenting our main result, we discuss some related works.  Our
work is motivated by \cite{Bose03,Subrahmanyam03} that study
quantum state transfer and entanglement distribution in a spin chain
of $N$ spins under the {\em free} evolution according to the
Hamiltonian
\begin{equation}\label{H1}
H_G =  - \frac{J}{2}\sum\limits_{\left\langle {i,j} \right\rangle
} {} \vec \sigma_i  \cdot \vec \sigma_j  - B\sum\limits_i {}
\sigma _{zi},
\end{equation}
where $\vec \sigma_i  \cdot \vec \sigma_j \equiv
\sigma_{\!x\,i} \, \sigma_{\!x\,j} +
\sigma_{\!y\,i} \, \sigma_{\!y\,j} +
\sigma_{\!z\,i} \, \sigma_{\!z\,j}$.
Perfect (near perfect) state transfer between the remotest spins are
only allowed for a 4-spin closed (open) chain \cite{Bose03}, apart
from the trivial case of 2 spins.

\ps

Our work is also motivated by \cite{Cubitt03}, which considers a
similar $3$-spin chain with only nearest neighbor coupling.  They
found mixed separable initial states such that the two remote spins
become entangled by the Hamiltonian without ever entangling the
mediating spin.  They also proved that such phenomenon is impossible
for pure states.  In our system, the mediating spin is entangled
during the process, but is completely disentangled from the data
qubits at the end, leaving the data qubits maximally entangled.

\ps

Our work can also find application in the quantum computation scheme
in \cite{Benjamin03}.  Reference \cite{Benjamin03} investigated
effective gates between non-neighboring spins in a Heisenberg spin
chain (\ref{H1}) with untunable couplings.  Our result is an analog in
the XY model (\ref{Hgen}), and has the merit of producing a much
simpler two-qubit gate (the composition of a swap gate and a
controlled-phase gate) compared to \cite{Benjamin03}.

\ps

Time optimal construction of indirect $2$-qubit gates, such as {\sc
swap} or {\sc cnot}, is considered in \cite{Khaneja03}. Our approach
is different in requiring no local manipulations, and also in
obtaining an exact implementation of the gate.

\ps

During the initial preparation of our manuscript, related works
\cite{Christandl03} were reported independently by Christandl {\it et
al}.  They considered perfect quantum state transfer over more general
spin networks (besides linear chains) under the same Hamiltonian
\eq{Hgen}.
When the interaction strength is homogeneous, i.e. $\omega_{ij} =
\omega$, they found that perfect state transfer is possible only
between the two ends of an $N=2$ or $3$ linear chain, or between the
antipoles for a hypercube.
When the coupling strength can be suitably tuned, perfect state
transfer is possible for any linear chain.
Extensions to a Heisenberg chain \eq{H1} was considered when a
tunable, local magnetic field can be applied to each spin.
Our motivating scheme in Sec.~\ref{sec:3pins-motivate} coincides with
the $N=3$ case of \cite{Christandl03}, which also motivated some
extensions in Sec.~\ref{sec:largen} for longer chains.
Our work differs in explicitly considering the evolution on all three
spins (instead of a subspace relevant to quantum state transfer as
considered in \cite{Christandl03}).
We exploit different initial configurations that result in more
flexibility and better efficiencies in tasks other than state
transfer, beyond immediate implications of the ability to perform
state transfer.

\vspace*{1ex}

\section{An effective two-qubit gate}

\subsection{Motivating scheme}
\label{sec:3pins-motivate}

The Hamiltonian for our $3$-spin chain \eq{Hgen} is simply
\begin{equation}\label{H2}
\!\!\!\!
H = \omega \left( {\sigma _ {+1} \, \sigma _ {-2}  + \sigma
_ {-1} \, \sigma _ {+2} } \right) + \lambda \left( {\sigma
_ {+2} \, \sigma _ {-3}  + \sigma _ {-2} \, \sigma _ {+3} } \right).
\end{equation}
We first motivate our result by considering quantum state transfer,
say, from the first to the last spin:
\begin{equation}\label{task}
\left| \varphi  \right\rangle _1 \left| 0 \right\rangle _2 \left|
0 \right\rangle _3  \to \left| 0 \right\rangle _1 \left| 0
\right\rangle _2 \left| \varphi  \right\rangle _3.
\end{equation}
Suppose the initial state of the first spin is $\left| \varphi
\right\rangle _1 = \alpha \left| 0 \right\rangle _1  + \beta
\left| 1 \right\rangle _1$, where $\left| \alpha  \right|^2  +
\left| \beta  \right|^2  = 1$.  The initial state of the three
spins can be written in the form,
\begin{eqnarray}
\left| {\psi \left( 0 \right)} \right\rangle  &=& \left( {\alpha
\left| 0 \right\rangle _1  + \beta \left| 1 \right\rangle _1 }
\right)\left| 0 \right\rangle _2 \left| 0 \right\rangle _3  \\
&=& \alpha \left| {000} \right\rangle  + \beta \left| {100}
\right\rangle. \nonumber
\end{eqnarray}
The state will evolve, under the Hamiltonian (\ref{H2}), as
\begin{equation}\label{psit}
\left| {\psi \left( t \right)} \right\rangle  = \alpha \left|
{000} \right\rangle  + \beta \exp \left( { - iHt} \right)\left|
{100} \right\rangle,
\end{equation}
where the first term is not affected by the Hamiltonian because
$H\left| {000} \right\rangle = 0$. The second term can be
evaluated analytically.  To do this, we make use of the special
properties of the raising and lowering operators to obtain
\begin{equation}\label{annihilation}
\begin{array}{l}
 H\left| {100} \right\rangle  = \omega \left| {010} \right\rangle,  \\
 H\left| {010} \right\rangle  = \omega \left| {100} \right\rangle  + \lambda \left| {001} \right\rangle,  \\
 H\left| {001} \right\rangle  = \lambda \left| {010} \right\rangle.  \\
 \end{array}
\end{equation}
Using \eq{annihilation}, the Taylor series expansion can be evaluated
explicitly
\begin{equation}
\begin{array}{l}
  \exp \left( { - iHt} \right)\left| {100} \right\rangle
= \left({1 +
       \frac{{\w^2 }}{{\g^2 }} \left( {\cos \g t - 1} \right)} \right)
       \left|{100} \right\rangle
\\
  ~~~~+\left( {\frac{{\w \lambda }}{{\g^2 }}
       \left({\cos \g t -1} \right)} \right)\left| {001} \right\rangle
     - i\frac{\omega }{\gamma }\sin \gamma t\left| {010} \right\rangle,  \\
 \end{array}
\end{equation}
where $\gamma  = \sqrt {\omega ^2  + \lambda ^2 }$. When $\lambda
= \omega$ and $t = \tau \equiv \smfrac{\pi}{\sqrt 2 \omega}$, we
have
\begin{equation} \label{exp}
\exp \left( { - iH\tau } \right)\left| {100} \right\rangle  =  -
\left| {001} \right\rangle,
\end{equation}
and
\begin{eqnarray}
 \left| {\psi \left( \tau  \right)} \right\rangle  &=& \alpha \left| {000} \right\rangle  - \beta \left| {100} \right\rangle  \\
  &=& \left| 0 \right\rangle _1 \left| 0 \right\rangle _2 \left( {\alpha \left| 0 \right\rangle _3  - \beta \left| 1 \right\rangle _3 }
  \right). \nonumber
\end{eqnarray}
Perfect state transfer \eq{task} is then obtained by applying the local
unitary operator $\sigma_{\!z\,3}$ to the last spin.

\subsection{The effective two-qubit gate}

To obtain our key result and to better understand the above
scheme, we examine the full matrix representation of the unitary
operator $U \equiv \exp(-i H \tau)$, with the basis ordered as
$|000\>$, $|001\>$, $|100\>$, $|101\>$, $|010\>$, $|011\>$,
$|110\>$, $|111\>$,
\bea \label{Ugate}
U & = & \left( \begin{array}{cccccccc}
    1 &  0 &  0 &  0 &  0 &  0 &  0 &  0
\\  0 &  0 & -1 &  0 &  0 &  0 &  0 &  0
\\  0 & -1 &  0 &  0 &  0 &  0 &  0 &  0
\\  0 &  0 &  0 & -1 &  0 &  0 &  0 &  0
\\  0 &  0 &  0 &  0 & -1 &  0 &  0 &  0
\\  0 &  0 &  0 &  0 &  0 &  0 & -1 &  0
\\  0 &  0 &  0 &  0 &  0 & -1 &  0 &  0
\\  0 &  0 &  0 &  0 &  0 &  0 &  0 &  1
\end{array} \right).
\eea
The subspaces spanned by $|0\>_2$ and $|1\>_2$ are {\em invariant}
under $U$. Thus restricting $U$ to either subspace forms an
effective two-qubit gate for the remote spins $1$ and $3$.
In particular, we can consider the restriction $\tilde{U}$ of $U$ to
the ``$|0\>_2$ subspace,''
\begin{equation}\label{Umatrix}
\tilde{U}_{13} = \left( {\begin{array}{*{20}c}
   1 & 0 & 0 & 0  \\
   0 & 0 & { - 1} & 0  \\
   0 & { - 1} & 0 & 0  \\
   0 & 0 & 0 & { - 1}  \\
\end{array}} \right),
\end{equation}
which is a composition of the {\sc swap} gate and a
``joint-phase-gate'' Diag$(1,-1,-1,-1)$ that is equivalent to the
controlled-$\s_z$ acting on spins $1$ and $3$ (up to renaming of basis
states and an overall sign).

\ps

In the scheme discussed in Sec.~\ref{sec:3pins-motivate}, spin $3$
is initialized as $|0\>_3$; the effect of the joint-phase-gate can
be reverted by the {\em local} $\s_{\!z\,3}$, so that the net
effect is to swap spins $1$ and $3$, resulting in perfect state
transfer.

\subsection{State transfer using $\tilde{U}_{13}$}
\label{sec:coherent}

Equation (\ref{Umatrix}) shows that perfect state transfer (up to an
irrelevant overall ``$-$'' sign) can be performed without the final
local gate $\s_{\!z\,3}$, if spin $2$ and spin $3$ are initialized to
be $|0\>_2|1\>_3$ instead of $|0\>_2|0\>_3$.  (Alternatively, we can
also initialize them to be $|1\>_2|0\>_3$.)  Thus, barring
initialization of spins $2$ and $3$, no local manipulation is required
at all!  As a final remark, state transfer due to \eq{Umatrix} is
coherent, thus it is {\em entanglement-preserving} and in particular,
$ \left| {\phi} \right\rangle_{a1} |01\>_{23} \rightarrow -\left| 10
\right\rangle_{12} \left| {\phi} \right\rangle _{a3}$ for any joint
state $\left| {\phi} \right\rangle_{a1}$ on spin 1 and an ancillary
system $a$ of arbitrary dimension.


\vspace*{1ex}

\subsection{Classical communication and entanglement generation}

The ability to transfer a certain amount of quantum information from
one end of a spin chain to another obviously implies the ability to
transfer at least as much classical communication, and to {\em share}
at least as much entanglement (in units of ebit) between the two ends
when the transfer is entanglement preserving.  In this section, we
present communication and entanglement sharing or generation schemes
(they differ by whether entanglement is distributed or created) that
outperform the above obvious methods.

\ps

Equation (\ref{Umatrix}) suggests a simple method for bidirectional
classical communication~\cite{BHLS03} between spins $1$ and $3$.
Suppose spins $1$ and $3$ are in the possession of Alice and Bob
respectively.  Spin $2$ is in the control of the communication company
who initializes the spin to $|0\>_2$ but does nothing otherwise.
Since $\tilde{U}$ acts exactly as a {\sc swap} gate on classical
states, Alice and Bob can each send one classical bit to the other.
Note that the combined communication rate is twice the rate of quantum
state transfer and that the qualitative ability to communicate
simultaneously in both direction is not implied from the ability
to perform quantum state transfer.

\ps

Equation (\ref{Umatrix}) also suggests an obvious method for creating
entanglement between spins $1$ and $3$ -- the gate $\tilde{U}$ creates
one ebit starting from the initial product state $|+\>_1 |+\>_3$,
where $|\pm\> = \smfrac{1}{\sqrt{2}} (|0\> \pm |1\>)$. As $\tilde{U}$
acts like the {\sc swap} gate followed by the joint-phase-gate, the
{\sc swap} leaves the state invariant, and the joint-phase-gate takes
$|+\>_1 |+\>_3$ to $\smfrac{1}{\sqrt{2}} (|0\>_1 |-\>_3 - |1\>_1
|+\>_3)$, which is maximally entangled.  Spin $2$ is invariant under
$U$ and is disentangled from spins $1$ and $3$ at the end.  Finally,
$\tilde{U}$ can be used to {\em share} two ebits if applied to an
initial state $|\Phi\>_{a1} \otimes |\Phi\>_{3b}$ where $|\Phi\> =
\smfrac{1}{\sqrt{2}}(|00\>+|11\>)$ and $1,a$ and $3,b$ are possessed
by Alice and Bob respectively.

\ps

The above methods are optimal if $\tilde{U}$ is a fixed given resource
and they require no local control besides local state initialization.
Exactly one of the {\sc swap} and the phase-gate components of
$\tilde{U}$ are used in each protocol.

\ps

More generally, given the Hamiltonian $H$, there are alternative
schemes for the various tasks that do not involve implementing
$\tilde{U}$ and are more efficient.  For example, let $\sqrt{U} \equiv
\exp(-iH\smfrac{\tau}{2})$, which can be written explicitly (again in
the basis $|000\>$, $|001\>$, $|100\>$, $|101\>$, $|010\>$, $|011\>$,
$|110\>$, $|111\>$) as
\bea
\sqrt{U} = \left(
\begin{array}{cccccccc}
   1 &     0 &     0 &     0 &     0 &     0 &     0 &     0 \\
   0 &   \p5 &  -\p5 &     0 & \sq12 &     0 &     0 &     0 \\
   0 &  -\p5 &   \p5 &     0 & \sq12 &     0 &     0 &     0 \\
   0 &     0 &     0 &     0 &     0 & \sq12 & \sq12 &     0 \\
   0 & \sq12 & \sq12 &     0 &     0 &     0 &     0 &     0 \\
   0 &     0 &     0 &     \sq12 & 0 &   \p5 &  -\p5 &     0 \\
   0 &     0 &     0 &     \sq12 & 0 &  -\p5 &   \p5 &     0 \\
   0 &     0 &     0 &     0 &     0 &     0 &     0 &     1
\end{array}
\right) .
\eea
The fourth and the fifth columns imply that
\begin{eqnarray}
\sqrt{U} \; |0\>_2 \left| {11} \right\rangle_{13} = -i |1\>_2 \;
\smfrac{1}{{\sqrt 2 }}\left( {\left| {01} \right\rangle  + \left|
{10} \right\rangle }
\right)_{13}, \label{first} \\
\sqrt{U}\;  |1\>_2 \left| {00} \right\rangle_{13} = -i |0\>_2 \;
\smfrac{1}{{\sqrt 2 }}\left( {\left| {10} \right\rangle  + \left|
{01} \right\rangle } \right)_{13}. \label{second}
\end{eqnarray}
Applying $H$ for a period of time $\smfrac{\tau}{2}$, half of that
required for implementing $\tilde{U}$, one can create a maximally
entangled state from a product state, without ancillas, and with spin
$2$ disentangled at the end.

\ps

More amusingly, Alice and Bob can repeat the procedure indefinitely
without resetting the mediating spin: Let spin $2$ be in $|0\>_2$
initially.  Alice and Bob input the states $|1\>_1 |1\>_3$, apply $H$
for time $\smfrac{\tau}{2}$, extract the maximally entangled state
(using \eq{first}), and then input the state $|0\>_1 |0\>_3$ instead.
By now, spin $2$ has evolved to $|1\>_2$, and according to
\eq{second}, another maximally entangled state can be created after
time $\smfrac{\tau}{2}$, and so on.  This requires only one
initialization of the mediating spin and the ability to prepare $|0\>$
and $|1\>$ locally by Alice and Bob, and to input/output a quantum
state of the data qubits at intervals separated by $\smfrac{\tau}{2}$.

\subsection{Effective two-qubit gate in a network}

\begin{figure}[http]
\begin{center}
\includegraphics[width=7 cm]{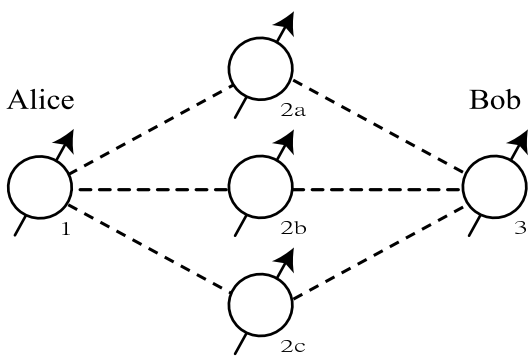}
\caption{A network of spins consists of three spin chains in
Fig.~\ref{3spins}.  An effective two-qubit gate, $\tilde U_{13}$, can
be obtained even if the coupling strength for the chains are
different.}
\label{3spins2}
\end{center}
\end{figure}

With the spin chain in Fig.~\ref{3spins} as a basic unit, an effective
two-qubit gate can be obtained in a larger network of spins. For
example, consider a network (see Fig.~\ref{3spins2}) with three
different homogeneous chains, evolved by a Hamiltonian $H_3 = H_a +
H_b + H_c$ where each of $H_{a,b,c}$ acts on one chain and has the
form of (\ref{H2}), and $\omega_a = \lambda_a$, $\omega_b = \lambda_b$,
and $\omega_c = \lambda_c$, but $\omega_a$, $\omega_b$ and $\omega_c$
can be different.
Let $\left| {\tilde 0} \right\rangle_2 \equiv \left| 0 \right\rangle
_{2a} \left| 0 \right\rangle _{2b} \left| 0 \right\rangle _{2c}$, and
$\left| {\tilde 1} \right\rangle_2 \equiv \left( {\omega _a \left| 1
\right\rangle _{2a} + \omega _b \left| 1 \right\rangle _{2b} + \omega
_c \left| 1 \right\rangle _{2c} } \right)/\omega$, where we re-define
$\omega \equiv (\omega _a^2 + \omega _b^2 + \omega_c^2 )^{1/2}$.  The
joint evolution is then analogous to the single chain case, and an
effective two-qubit gate $\tilde{U}_{13}$ in (\ref{Umatrix}) can be
obtained at $t=\tau$.  Similar argument applies to any number of
chains.  Each extra chain simply increases the coupling strength.
We note that similar techniques has been used in the context
of quantum walk, see for example \cite{Childs01}.

\section{Spin chains with $N>3$ spins}
\label{sec:largen}

In this section, we extend the above results to spin chains with $N>3$
spins. We first consider generating or sharing entanglement in
inhomogeneous chains and then consider quantum state transfer for
homogeneous chains.

\subsection{Inhomogeneous chains}

We first consider inhomogeneous chains in which the coupling strength
$\omega_{j \hspace*{0.3ex} j\!+\!1}$ can be tuned for different pairs
of spins.  Our entanglement generating or sharing schemes borrow
techniques from Christandl {\it et al} \cite{Christandl03}.  Our
scheme takes about half of the time required for quantum state transfer.

Consider a chain of $N$ spins.  We want to prepare a maximally
entangled state $\sh \left( {\left| 0 \right\rangle _1 \left| 1
\right\rangle _N + \left| 1 \right\rangle _1 \left| 0 \right\rangle _N
} \right)$ between the data spins $1$ and $N$ at the two ends.  When
$N$ is odd, we find an entanglement generating scheme with a simple
initial spin configuration.  When $N$ is even, we find an entanglement
sharing scheme that requires preexisting entanglement in the chain
and a local magnetic field. We are going to discuss these two cases
separately.

\subsubsection{Odd number of spins}

We use a more convenient basis defined as $\equiv \sh \left( {\left| {100
\cdots} \right\rangle + \left| {\cdots001} \right\rangle } \right)$,
$\left| {\tilde 2} \right\rangle \equiv \sh \left( {\left| {010\cdots}
\right\rangle + \left| {\cdots010} \right\rangle } \right)$, and
similarly for $|\tilde{j}\>$, $j=1,\cdots,n\!-\!1$, and $\left|
{\tilde n} \right\rangle \equiv \left| {0\cdots010\cdots0}
\right\rangle $.  Here, $n=\smfrac{1}{2}(N\!+\!1)$ is the position of
the middle spin.  Suppose the initial state is $\left| {\tilde n}
\right\rangle$.  Then, Hamiltonian in (\ref{Hgenv2}) acts in the
basis $\left| {\tilde 1} \right\rangle, \left| {\tilde 2} \right\rangle,
\cdots,\left| {\tilde n} \right\rangle $ as
\begin{equation}
H_{\rm odd}  = \left( {\begin{array}{*{20}c}\label{odd}
   0 & {\omega _{12} } & 0 &  \cdots  & 0  \\
   {\omega _{12} } & 0 & {\omega _{23} } &  \cdots  & 0  \\
   0 & {\omega _{23} } & 0 &  \cdots  &  \vdots   \\
    \vdots & \vdots & \vdots & \ddots & {\sqrt 2 \, \omega_{n\!-\!1\,n}} \\
   0 & 0 &  \cdots  & {\sqrt 2 \, \omega _{n\!-\!1\,n} } & 0  \\
\end{array}} \right).
\end{equation}
The task to generate entanglement is the same as to rotate $\left|
{\tilde n} \right\rangle$ into $\left| {\tilde 1} \right\rangle$, and
it is equivalent to transferring a quantum state in a chain of length
$n=\smfrac{1}{2}(N\!+\!1)$ as solved by Christandl {\it et al}
\cite{Christandl03}.  When the matrix elements $\omega _{j,j \!+\!
1}$ are chosen such that
\begin{equation}
\omega _{j,j + 1}  \propto \left\{ {\begin{array}{*{20}c}
   {\sqrt {j\left( {n - j} \right)} },  \\
   {\sqrt {n - 1} /\sqrt 2 },  \\
\end{array}} \right.\quad \begin{array}{*{20}c}
   {\rm for}~~{j \ne n - 1},  \\
   {\rm for}~~{j = n - 1},  \\
\end{array}
\end{equation}
\begin{equation}
\left| {\left\langle {\tilde 1} \right|\exp \left( {-i \, H_{\rm odd} \,t}
\right)\left| {\tilde n} \right\rangle } \right| = 1
\end{equation}
when $t =\pi /\lambda$ for a constant $\lambda$ described in
\cite{Christandl03}.
This concludes the entanglement generation scheme.

\ps

Note that the time for the above scheme is constant, while the
coupling strengths $\omega_{j,j + 1}$ is a function of $n$ (and most
are roughly proportional to $n$).  The resource in this framework can
be identified as $\omega_{j,j + 1} t$ or simply $\omega_{j,j + 1}$.
Thus, our scheme creates entanglement using roughly half of the
resource needed in the obvious method of entanglement sharing via quantum
state transfer.

\subsubsection{Even number of spins}

We describe a similar scheme for an even number of spins, which
requires an initially entangled state, and a time independent local
magnetic field.  (It is thus an entanglement sharing scheme.)
Let $|\tilde{j}\>$ be as defined before, except now $\left|
{\tilde n} \right\rangle \equiv \sh \left( {\left| {0\cdots10\cdots0}
\right\rangle + \left| {0\cdots01\cdots0} \right\rangle } \right)$.
Again, we initialize our state to $\left| {\tilde n} \right\rangle$,
and entanglement sharing is achieved by rotating $\left| {\tilde n}
\right\rangle$ to $\left| {\tilde 1} \right\rangle$.
The nontrivial part of the Hamiltonian in this basis is
\begin{equation}
H_{\rm even}  = \left( {\begin{array}{*{20}c}
   0 & {\omega _{12} } & 0 &  \cdots  & 0  \\
   {\omega _{12} } & 0 & {\omega _{23} } &  \cdots  & 0  \\
   0 & {\omega _{23} } & 0 &  \cdots  & 0  \\
    \vdots  &  \vdots  &  \vdots  &  \ddots  & {\omega _{n\!-\!1\,n} }  \\
   0 & 0 &  \cdots  & {\omega _{n\!-\!1\,n} } & {\omega _{n~n+1} }  \\
\end{array}} \right).
\end{equation}
$H_{\rm even}$ differs from $H_{\rm odd}$ only in the extra term
${\omega _{n,n + 1}}$ in the lower right corner.  The extra term can
be removed by applying a static magnetic field of strength
${\textstyle{1 \over 2}} \, \omega _{n,n + 1}$ to the middle two
spins.  Mathematically, this adds a local Hamiltonian $H_B = -
{\textstyle{1 \over 2}}\omega _{n,n + 1} \left( {\sigma _{z,n} +
\sigma _{z,n + 1} } \right)$, and since
%
%
$H_B |\tilde{n}\> = 0$,
$H_B |\tilde{j}\> = {\omega _{n,n + 1}} |\tilde{j}\>$
for all $j<n$, the diagonal terms in
$H_{\rm even} + H_B$ are all equal to ${\omega _{n,n + 1}}$ and
contribute an irrelevant phase only.
The rest follows from the odd case.

\ps

As a side remark, for a homogeneous chain when the coupling constants
$\omega _{ij}$ are equal, an even chain entangles slightly better.
The two extra factors of ${\sqrt 2 }$ in $H_{\rm odd}$ causes perfect
entanglement to be possible only for $N=3$, while it is possible
for $N=4$ and $6$ in the even case.

\subsection{Homogeneous chains}

For homogeneous spin chains, we study the fidelity of quantum state
transfer
\begin{equation}\label{taskn}
\left| \varphi \right\rangle _1 \left| 0 \right\rangle _2 \cdots
\left| 0 \right\rangle _N \to \left| 0 \right\rangle _1 \left| 0
\right\rangle _2 \cdots \left| \varphi \right\rangle _N.
\end{equation}
where the fidelity (averaged over inputs) \cite{Bose03} is defined as
\bea
\label{fidelity}
F & = & \frac{{{f\left( t \right)}}}{3} + \frac{{ {f\left( t
\right)}^2 }}{6} + \frac{1}{2} \,, {\rm ~~~where}
\\
f\left( t \right) & \equiv & | \left\langle 10\cdots0 \right|\exp
\left( {-iHt} \right)\left| 0\cdots01 \right\rangle |
\eea
and a maximization over $t$ is then taken.

\ps

\begin{figure}[http]
\begin{center}
\psfig{file=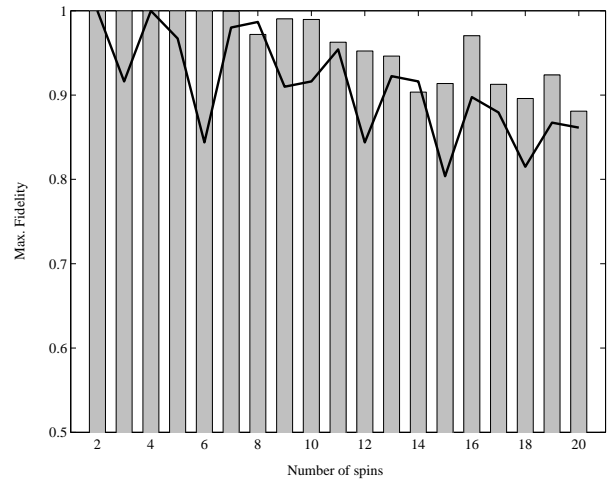,width=8cm}
\end{center}
\vspace*{-2ex} \caption{Average fidelity $f(t)$, maximized over ${t
\in (0,2000/\omega)}$ in (\ref{fidelity}), versus the number of spins
$N$ in the chain.  The bar chart and the line are for the XY \eq{Hgen}
and Heisenberg \eq{H1} chains respectively.}
\label{plotN}
\end{figure}

In Fig.~\ref{plotN}, we plotted numerical maximization of $f(t)$ over
a sufficient long but finite range of $t$.  A comparison with similar
numerical maximization for the Heisenberg chain (with Hamiltonian
given by \eq{H1}) suggests that the XY chain (with Hamiltonian given
by \eq{Hgen}) is more effective in quantum state transfer.

\ps

For asymptotically large $N$, using an expression for $f(t)$ in
Ref.~\cite{Christandl03} for the XY chain,
\begin{equation}
{{f\left( t \right)}}= \left| \frac{2}{N+1}\sum_{m=1}^N \sin{\frac{\pi
m}{N+1}}\sin{\frac{\pi m N}{N+1}} e^{-i E_m t} \right| \,,
\label{ftfromref6}
\end{equation}
where $E_m= - 2 \omega \cos{\frac{m\pi}{N+1}}$.  Let $J_k$ be the
Bessel functions.  The Jacobi-Anger expansion \cite{Arfken,mathca} states
that
\begin{eqnarray}
e^{iz\cos \theta}=\sum_{k=-\infty}^{\infty} i^k J_k(z) \; e^{-i k \theta} \,.
\end{eqnarray}
Applying this to \eq{ftfromref6}, taking $N$ to be large and $t_0$ to
be of order $N$, and neglecting the terms with $k \gg N$ or $k \ll N$,
we have,
\begin{equation}
f(t_0)\approx 2 \left| J_N(2\omega t_0)+J_{N+2}(2\omega t_0) \right| \,.
\end{equation}
Since $N+2+\xi(N+2)^{1/3} \approx N+\xi N^{1/3}$ for all $\xi$, and
$J_N(N+\xi N^{1/3})\approx(2/N)^{1/3} A(-2^{1/3}\xi)$
\cite{abramowitz}, where $A(\cdot)$ is the Airy function,
applying $2 \omega t_0= N+0.8089 N^{1/3}$ gives a lower bound on
the fidelity
\begin{equation}
f(t_0) \approx 2.6998 N^{-1/3} \,. \label{XY}
\end{equation}
This is twice the value of $f(t_0)$ for a Heisenberg chain
\eq{H1} for the same $t_0$ \cite{Bose03}.

\ps

\raggedbottom

\begin{figure}[http]
\begin{center}
\includegraphics[width=8.2cm]{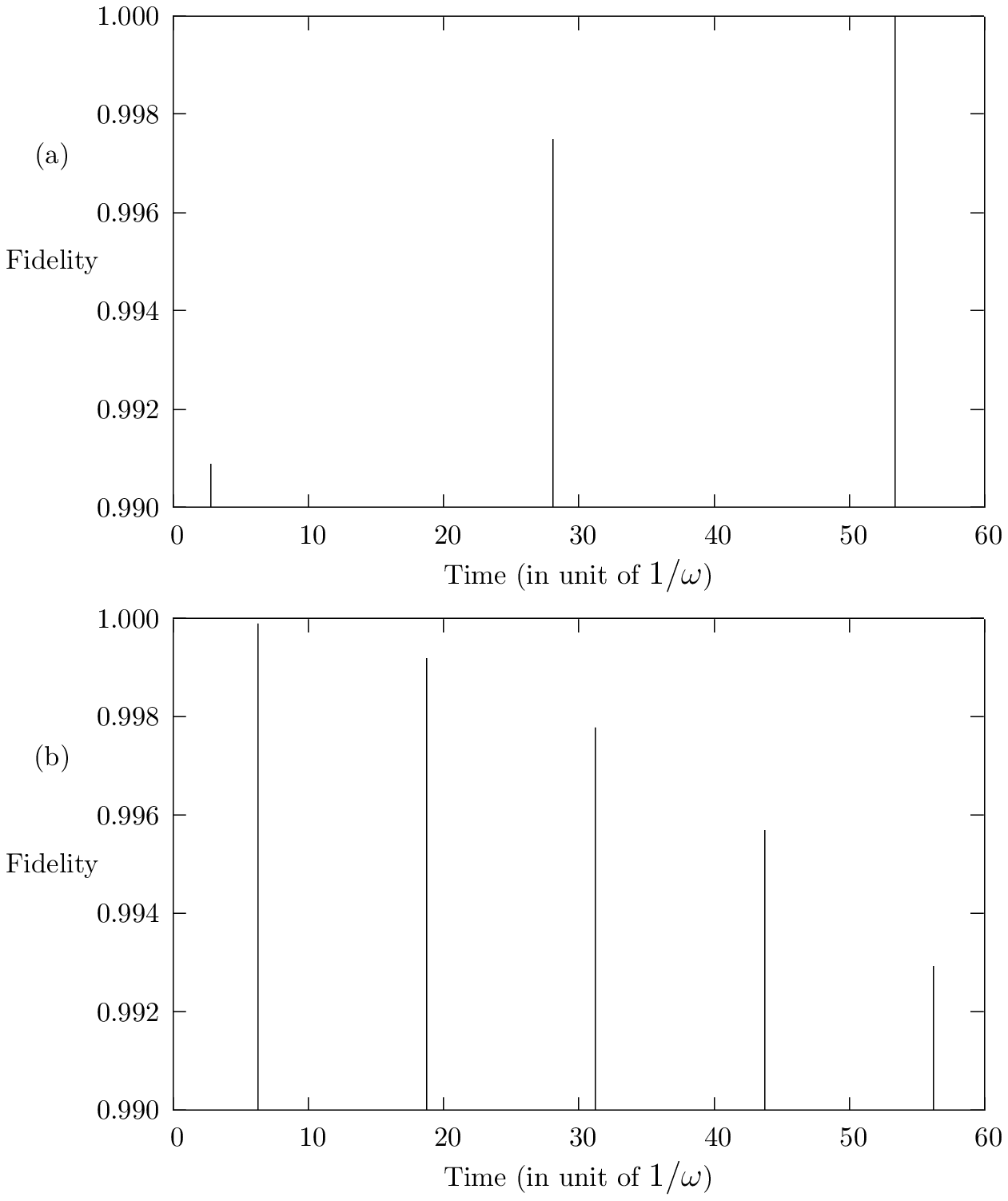}
\caption{Peaks of the average
fidelity $F(t)$ (\ref{fidelity}) as a function of time for an
$N=4$ spin chain.  Plot (a) represents the case without a magnetic
field, plot (b) represents the case when the middle two spins are
subject to a magnetic field of strength $B=0.625\omega$.  The
maximum peak is 0.99991 at $t=6.25/\omega$.} \label{fidelityB}
\end{center}
\end{figure}

Although perfect state transfer is possible only for $N=2$ and $N=3$,
fidelities very close to 1 are obtainable for sufficiently long
evolution times.  Such long evolution times can be shortened using a
non-homogeneous magnetic field applied to individual spins. For
example, for an $N=4$ XY chain, the maximum fidelity obtained is
$0.99997$ at $t=53.4/\omega$, in the time interval of
($0,100/\omega$). However, if we apply a magnetic field of strength
$B=0.625\omega$ on the two middle spins only, the fidelity is
$0.99991$ at $t=6.25/\omega$, see Fig.~\ref{fidelityB}.

\section{Conclusion}

We have demonstrated how to obtain an exact effective two-qubit gate
in a chain of three spins without local manipulations, and discussed
various applications of it, including quantum state transfer,
classical communication, and entanglement generation and
distribution. We also discuss some extensions to longer chains.  Other
observations of the system can be found in the Appendix.

\begin{acknowledgments}
We thank Andrew Landahl for help comments on an earlier manuscript.
This work is partially supported from the US National Science
Foundation under grant no. EIA-0086038.  DWL thanks the Richard C.
Tolman Endowment Fund at Caltech for partial support. SB thanks for a
postdoctoral scholarship at the Caltech Institute for Quantum
Information, during which this work started.
\end{acknowledgments}

\appendix
\section*{Appendix: Miscellaneous results}
\label{sec:others}
Here we explore more applications of the gate $U$ and other
results. In a chain of three spins, the gate $U$ can actually entangle
any pair of spins. For example, to entangle spin 1 and 2, while
leaving spin 3 unentangled, we can initialize the state to be $\left|
0 \right\rangle _1 \left| + \right\rangle _2 \left| + \right\rangle
_3$. The final state is $\frac{1}{{\sqrt 2 }}\left( {\left| 0
\right\rangle _1 \left| - \right\rangle _2 - \left| 1 \right\rangle _1
\left| + \right\rangle _2 } \right)\left| 0 \right\rangle _3$. On the
other hand, the gate $U$ can also act as an effective one-qubit gate,
$\sigma _{z2}$, for the middle spin if we initialize the state to be $
\left| 0 \right\rangle _1 \left( {\alpha \left| 0 \right\rangle _2 +
\beta \left| 1 \right\rangle _2 } \right)\left| 0 \right\rangle _3$,
for any $\alpha$ and $\beta$.

\ps

Together with an application of a single-qubit rotation, we can
readily generate a three-qubit entangled state, the $W$ state
\cite{DVC00}, $\frac{1}{{\sqrt 3 }}\left( {\left| {101} \right\rangle
+ \left| {011} \right\rangle + \left| {110} \right\rangle } \right) $,
by a free evolution under the Hamiltonian in (\ref{H2}).  More
explicitly, $\exp \left( { - iHt} \right)\left| {101} \right\rangle =
\cos \gamma t\left| {101} \right\rangle - i\frac{\omega }{\gamma }\sin
\gamma t\left| {011} \right\rangle - i\frac{\lambda }{\gamma }\sin
\gamma t\left| {110} \right\rangle$.  When $\omega=\lambda$ and $t =
\tan ^{ - 1} \left( {\sqrt 2 } \right)/\sqrt 2 \omega$, we obtain the
$W$ state after applying the phase gate $\left| 0 \right\rangle
\left\langle 0 \right| + i\left| 1 \right\rangle \left\langle 1
\right|$ to the middle spin.



\end{document}